\documentclass[twocolumn,aps,prl,floats,showpacs]{revtex4}%
\usepackage{epsfig}
\usepackage{dcolumn}
\usepackage{bm}
\usepackage{amsmath}
\usepackage{amsfonts}
\usepackage{amssymb}
\usepackage{graphicx}%
\begin{document}
\title{Glassy behavior of electrons near metal-insulator transitions}
\author{V. Dobrosavljevi\'{c}, D. Tanaskovi\'c, and A.A. Pastor}
\affiliation{Department of Physics and National High Magnetic Field Laboratory \linebreak
Florida State University, Tallahassee, Florida 32306}
\date{\today}

\begin{abstract}
The emergence of glassy behavior of electrons is investigated for systems
close to the disorder and/or interaction-driven metal-insulator transitions.
Our results indicate that Anderson localization effects strongly stabilize
such glassy behavior, while Mott localization tends to suppress it. We predict
the emergence of an intermediate metallic glassy phase separating the
insulator from the normal metal. This effect is expected to be most pronounced
for sufficiently disordered systems, in agreement with recent experimental observations.

\end{abstract}
\maketitle

In many disordered electronic \cite{lr} systems, electron-electron
interactions and disorder are equally important, and lead to a rich variety of
behaviors which remain difficult do understand. Their competition often leads
to the emergence of many metastable states and the resulting history-dependent
glassy dynamics of electrons. Such behavior has often been observed in
sufficiently low mobility materials \cite{ovadyahu,zvi}, but more recent
experiments \cite{popovic,jan} have provided striking and precise information
of such non-ergodic dynamics.

Theoretically, the possibility for glassy behavior in the charge sector has
been anticipated a long time ago \cite{efros} in situations where the
electrons are strongly localized due to disorder. In the opposite limit, for
well delocalized electronic wavefunctions, one expects a single well defined
ground state and absence of glassiness. The behavior in the intermediate
region has \ proved more difficult to understand, and at present little is
known as to the precise role and stability of the glassy phase close to the
metal-insulator transition (MIT) \cite{mott}. \ On physical grounds, one
expects the quantum fluctuations \cite{pastor} associated with mobile
electrons \ to suppress glassy ordering, but their precise effects remain to
be elucidated. Note that even the \emph{amplitude }of such quantum
fluctuations must be a singular function of the distance to the MIT, since
they are dynamically determined by processes that control the electronic mobility.

To clarify the situation, the following basic questions need to be addressed:
(1) Does the MIT\ coincide with the onset of glassy behavior? (2) How do
different physical processes that can localize electrons \ affect the
stability of the glass phase? In the following, we provide simple and
physically transparent \ answers to both questions. We find that: (a) Glassy
behavior generally emerges before the electrons localize; (b) Anderson
localization \cite{anderson} enhances the stability of the glassy phase, while
Mott localization \cite{mott} tends to suppress it. We thus predict the
emergence of an intermediate metallic glass phase separating the ordinary
metal from the insulator. However, we expect this effect to be of relevance
only for sufficiently strong disorder, consistent with recent experiments
\cite{popovic}.

As a simplest model where both Anderson and Mott routes to localization can
occur, we consider an extended Hubbard model of disordered spin $1/2$
electrons, as given by the Hamiltonian%

\[
H=\sum_{ij\sigma}(-t_{ij}+\varepsilon_{i}\delta_{ij})c_{i,\sigma}^{\dagger
}c_{j,\sigma}+U\sum_{i}n_{i\uparrow}n_{i\downarrow}+\sum_{ij}V_{ij}\delta
n_{i}\delta n_{j}.
\]
Here, $\delta n_{i}=$ $n_{i}-$ $\left\langle n_{i}\right\rangle $ represent
local density fluctuations ( $\left\langle n_{i}\right\rangle $ is the
site-averaged electron density), $U$ is the on-site interaction, and
$\varepsilon_{i\ }$are$_{\ }$Gaussian distributed random site energies of
variance $W^{2}$. In order to allow for glassy freezing of electrons in the
charge sector, we introduce weak inter-site density-density interactions
$V_{ij}$, which we also also choose to be Gaussian distributed random
\cite{random} variables of variance $V^{2}$ /$z$ ($z$ is the coordination
number). We emphasize that, in contrast to previous work \cite{pastor}, we
shall keep the coordination number $z$ finite, in order to allow for the
possibility of Anderson localization. To investigate the emergence of glassy
ordering, we formally average over disorder by using standard replica methods
\cite{pastor}, and introduce collective $Q$-fields to decouple the inter-site
$V$-term \cite{pastor}. As a result, the replicated partition function can be
written in the form%
\begin{equation}
Z^{n}=\int%
{\displaystyle\prod\limits_{i}}
Dc_{i}^{\dagger}Dc_{i}DQ_{i}\,d\varepsilon_{i}P(\varepsilon_{i})\exp\{-S\,\},
\end{equation}
where $S=S_{el}+S_{Q}+S_{int},$ with%

\begin{align}
S_{el}=  &
{\displaystyle\sum_{ij\sigma}}
\sum_{a}%
{\displaystyle\int\limits_{0}^{\beta}}
d\tau\,c_{i\sigma}^{\dagger a}(\tau)\left[  (\partial_{\tau}-\mu
+\varepsilon_{j})\delta_{ij}-t_{ij}\right]  c_{j\sigma}^{a}(\tau)\nonumber\\
+  &  U\sum_{a}%
{\displaystyle\int\limits_{0}^{\beta}}
d\tau\,c_{i\uparrow}^{\dagger a}(\tau)c_{i\uparrow}^{a}(\tau)c_{i\downarrow
}^{\dagger a}(\tau)c_{i\downarrow}^{a}(\tau),
\end{align}

\begin{equation}
S_{Q}=\frac{1}{2}V^{2}%
{\displaystyle\sum\limits_{ij,ab}}
{\displaystyle\int\limits_{0}^{\beta}}
{\displaystyle\int\limits_{0}^{\beta}}
d\tau d\tau^{\prime}\,Q_{i}^{ab}(\tau-\tau^{\prime})K_{ij}Q_{j}^{ab}(\tau
-\tau^{\prime}),
\end{equation}
with%

\begin{equation}
S_{int}=-\frac{1}{2}V^{2}%
{\displaystyle\sum\limits_{i,ab}}
{\displaystyle\int\limits_{0}^{\beta}}
{\displaystyle\int\limits_{0}^{\beta}}
d\tau d\tau^{\prime}\,\delta n_{i}^{a}(\tau)Q_{i}^{ab}(\tau-\tau^{\prime
})\delta n_{i}^{b}(\tau^{\prime}).
\end{equation}

Here, $c_{j\sigma}^{a}(\omega_{m})$ are the replicated Grassmann fields
\cite{pastor} corresponding to electrons, $\delta n_{i}^{a}(\tau)=\sum
_{\sigma}[c_{i\sigma}^{\dagger a}(\tau)c_{i\sigma}^{a}(\tau)\,-\,<c_{i\sigma
}^{\dagger a}(\tau)c_{i\sigma}^{a}(\tau)>],$ $a=1,...n$ $(n\rightarrow0)$ are
the replica indices, and $K_{ij}=$ $\frac{2}{z}f_{ij}^{\,\,-1}$is the inverse
lattice matrix corresponding to the interactions $V_{ij\text{. }}$

\textit{Saddle-point theory. }To investigate the glassy behavior on a
mean-field level \cite{pastor}, we formally integrate out the conduction
electrons to produce an effective action $S_{eff}[Q]$ for the $Q$-fields, the
variation of which produces the following saddle-point conditions

\begin{align}
Q_{i}^{aa}(\tau-\tau^{\prime})  &  =\chi(\tau-\tau^{\prime})=\left\langle
\,\delta n_{i}^{a}(\tau)\delta n_{i}^{a}(\tau^{\prime})\right\rangle _{SP},\\
Q_{i}^{ab}(\tau-\tau^{\prime})  &  =q^{ab}=\left\langle \,\delta n_{i}%
^{a}(\tau)\delta n_{i}^{b}(\tau^{\prime})\right\rangle _{SP};(a\neq b).
\end{align}
Physically, $\chi(\tau-\tau^{\prime})$ is the averaged local dynamic
compressibility, and $q^{ab}$ is related to the familiar Edwards-Anderson
order parameter \cite{pastor}. In these expressions, the averages are taken
with respect to the saddle-point action for conduction electrons, as given by
\begin{equation}
S_{sp}=S_{el}[c_{i}^{\dagger},c]+S_{int}[c_{i}^{\dagger},c;\chi,q^{ab}]
\end{equation}

\textit{Glass transition. }In our approach, the emergence of many metastable
states corresponding to glassy ordering is identified as a replica symmetry
breaking (RSB) instability \cite{pastor}. To perform such a stability
analysis, we evaluate the $S_{eff}[Q]$ at the saddle point, and then examine
its variation to infinitesimal RSB\ perturbations of the form $q^{ab}=q+\delta
q^{ab}$. The corresponding stability matrix can be expressed through
appropriate correlation function of density fluctuations, and the rest of the
analysis is identical as in the classical case \cite{spinglass,pastor}. The
resulting instability criterion, which corresponds to the vanishing of the
relevant eigenvalue of the stability matrix, takes the form%

\begin{equation}
\lambda_{o}=1-V^{2}%
{\displaystyle\sum\limits_{j}}
[\chi_{ij}^{2}(\omega_{n}=0)]_{dis}=0. \label{instab}%
\end{equation}
Here, the non-local static compressibilities are defined (for a fixed
realization of disorder) as%

\begin{equation}
\chi_{ij}(\omega_{n}=0)=-\partial n_{i}/\partial\varepsilon_{j}%
\end{equation}
where $n_{i}$ is the local expectation value of the electron density, and
$[\cdots]_{dis}$ represents the average over disorder.

The eigenvalue $\lambda_{o}$ play the same role in our theory as the parameter
$r$ in a conventional Landau-Ginzburg action, the vanishing of which indicates
an ordering instability in the appropriate channel. Such parameters are
typically assumed to be smooth (regular) functions of control parameters such
as temperature or the Fermi energy. In our case the situation is more
interesting: we will show that the correlation function $\chi^{(2)}\equiv
\sum_{j}[\chi_{ij}^{2}]_{dis\text{ }}$is a \emph{singular }function,
\emph{diverging }at an Anderson-like metal-insulator transition. In this way,
Anderson localization can be regarded as a singular perturbation in the case
of the glassy ordering of electrons. We should emphasize that this unusual
sensitivity to Anderson localization is \emph{not }found in cases of more
conventional transitions to uniform ordering. Our analysis can be easily
repeated in such situations, and the corresponding instability criterion would
instead involve the average compressibility evaluated at the relevant ordering
wave vector $\overrightarrow{K}$, as given by%

\begin{equation}
\chi^{(1)}(\overrightarrow{K})=\sum_{j}e^{i\overrightarrow{K}\cdot
\overrightarrow{j}}[\chi_{ij}(\omega_{n}=0)]_{dis}.
\end{equation}
It is well known \cite{lr} that such quantities remain \emph{finite} (see also
below) and thus non-singular at an Anderson transition. Similar behavior is
found in the case of a superconducting instability in presence of localization
\cite{kotlkapitul}. In this case, the relevant pairing susceptibility (in the
Cooper channel) was found to remain noncritical at the Anderson transition,
opening the possibility for a direct superconductor-insulator transition, as
seen in many experiment.

For general values of disorder $W$, and interactions $U$ and $V$, evaluating
$[\chi_{ij}^{2}]_{dis\text{ }}$is difficult, since it has to be computed with
respect to the action $S_{sp}[c_{i}^{\dagger},c]$ describing disordered
interaction electrons in finite dimensions. The situation is simpler both in
limits of very strong and very weak disorder, where reliable approximations
are available. We first examine the limit of very strong disorder, and
determine the critical value of the Fermi energy corresponding to the
emergence of glassy ordering at $T=0$. To determine the transition line, to
leading order in $W/U$ and $W/V$ it is sufficient to set $U=V=0$ in computing
the required quantity $\chi^{(2)}$.

\textit{High disorder - Anderson transition. }As the disorder grows, the
system approaches the Anderson transition at $t=t_{c}(W)\sim W$. The first
hint of singular behavior of $\chi^{(2)}$ in an Anderson insulator is seen by
examining the deeply insulating, i. e. atomic limit $W\gg t,$ where to leading
order we set $t=0$ and obtain $\chi_{ij}=\delta(\varepsilon_{i}-\mu
)\delta_{ij}$. We immediately find that $\chi^{(1)}(\overrightarrow
{K})=[\delta(\varepsilon_{i}-\mu)]_{dis}=P(\varepsilon=\mu)$ remains finite,
but $\chi^{(2)}=[\delta^{2}(\varepsilon_{i}-\mu)]_{dis}=+\infty$ diverges!
Since we expect all quantities to behave in qualitatively the same fashion
throughout the insulating phase, we anticipate $\chi^{(2)}$ to diverge already
at the Anderson transition. Note that, since the instability of the glassy
phase occurs already at $\chi^{(2)}=V^{-2}$, the glass transition must
\emph{precede} the localization transition. Thus, for any finite inter-site
interaction $V$, we predict the emergence of an intermediate \emph{metallic
glass phase }separating the Fermi liquid from the Anderson insulator. Assuming
that near the transition%

\begin{equation}
\chi^{(2)}\simeq\frac{A}{W^{2}}((t/W)-B)^{-\alpha}%
\end{equation}
($A$ and $B=t_{c}/W$ are constants of order unity), from Eq. $\left(
\ref{instab}\right)  $ we can estimate the form of the glass transition line,
and we get
\begin{equation}
\delta t(W)=t_{G}(W)-t_{c}(W)\sim V^{2/\alpha}W^{1-2/\alpha};\;W\rightarrow
\infty\label{anderson}%
\end{equation}
The glass transition and the Anderson transition lines are predicted to
converge at large disorder for $\alpha<2,$ and diverge for $\alpha>2$. Since
all the known exponents characterizing the localization transition seem to
grow with dimensionality, we may expect a particularly large metallic glass
phase in large dimensions.

\textit{ }In order to confirm this scenario by explicit calculations, we
compute the behavior of $\chi^{(2)}$ at the Anderson transition of a
half-filled Bethe lattice of coordination $z=3.$We use an essentially exact
numerical approach \cite{motand} based on the recursive structure of the Bethe
lattice \cite{abou-chacra}. In this approach, local and non-local Green's
functions on a Bethe lattice can be sampled from a large ensemble, and the
compressibilities $\chi_{ij}$ can be then calculated by examining how a local
charge density $n_{i}$ is modified by an infinitesimal variation of the local
site energy $\varepsilon_{j}$ on another site. To do this, we have taken
special care in evaluating the local charge densities $n_{i}$ by numerically
computing the required frequency summations over the Matsubara axis, where the
numerical difficulties are minimized. Using this method, we have calculated
$\chi^{(2)}$as a function of $W/t_{\text{ }}$(for this lattice at half-filling
$E_{F\text{ }}=$ 2$\sqrt{2}t$ ), and find that it decreases exponentially
\cite{mirlin} as the Anderson transition is approached. We emphasize that only
a finite enhancement of $\chi^{(2)}$ is required to trigger the instability to
glassy ordering, which therefore occurs before the Anderson transition is
reached. The resulting $T=0$ phase diagram, valid in the limit of large
disorder, is presented in Fig. 1. Note that the glass transition line in this
case has the form $t_{G}(W)\sim W$, in agreement with the fact that
exponential critical behavior of $\chi^{(2)}$ corresponds to $\alpha
\rightarrow\infty$ in the above general scenario. These results are strikingly
different from those obtained in a theory which ignores localization
\cite{pastor}, where $t_{G}(W)$ was found to be weakly dependent on disorder,
and remain\emph{ finite} as $W\longrightarrow\infty$. Anderson localization
effects thus strongly enhance the stability of the glass phase at sufficiently
large disorder. Nevertheless, since the Fermi liquid to metallic glass (FMG)
transition occurs\emph{ }at a finite distance \emph{before} the localization
transition, we do not expect the leading quantum critical behavior
\cite{denis} at the FMG transition to be qualitatively modified by the
localization effects.%

\begin{figure}
[ptb]
\begin{center}
\includegraphics[
height=2.0232in,
width=2.919in
]%
{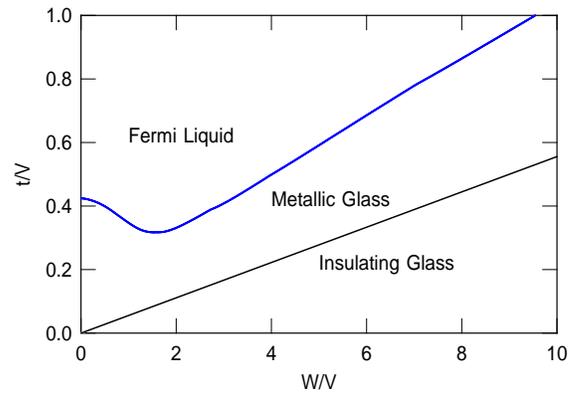}%
\caption{Phase diagram for electrons on the $z=3$ Bethe lattice, where
localization effects are treated using an exact numerical method. The results
are valid in the limit of large disorder ($W/U\rightarrow\infty$). }%
\label{fig1}%
\end{center}
\end{figure}

\textit{ Low disorder - Mott transition. }In the limit of weak disorder $W\ll
U,V$, and interactions drive the metal-insulator transition. Concentrating on
the model at half-filling, the system will undergo a Mott transition
\cite{mott} as the hopping $t$ is sufficiently reduced. Since for the Mott
transition $t_{Mott}(U)\sim U$, near the transition $W\ll t$, and to leading
order we can ignore the localization effects. In addition, we assume that
$V\ll U,$ and to leading order the compressibilities have to be calculated
with respect to the action $S_{el}$ of a disordered Hubbard model. The
simplest formulation that can describe the effects of weak disorder on such a
Mott transition is obtained from the dynamical mean-field theory (DMFT)
\cite{dmft}. This formulation, which ignores localization effects, is obtained
by rescaling the hopping elements $t\rightarrow t/\sqrt{z}$ and then formally
taking the limit of large coordination $z\rightarrow\infty$. To obtain
qualitatively correct analytical results describing the vicinity of the
disordered Mott transition at $T=0,$ we have solved the DMFT equations using a
4-boson method of Kotliar and Ruckenstein \cite{kotlruck}. At weak disorder,
these equations can be easily solved in close form, and we simply report the
relevant results. The critical value of hopping for the Mott transition is
found to decrease with disorder, as%
\begin{equation}
t_{c}(W)\approx t_{c}^{o}\,(1-4(W/U)^{2}+\cdots),
\end{equation}
where for a simple semi-circular density of states \cite{dmft} $t_{c}^{o}=3\pi
U/64$ (in this model, the bandwidth $B=4t$). Physically, the disorder tends to
suppress the Mott insulating state, since it broadens the Hubbard bands and
narrows the Mott-Hubbard gap. At sufficiently strong disorder $W\geq U$, the
Mott insulator is suppressed even in the atomic limit $t\rightarrow0$. The
behavior of the compressibilities can also be calculated near the Mott
transition, and to leading order we find%
\begin{equation}
\chi^{(2)}=\left[  \frac{8}{3\pi t_{c}^{o}}(1-\frac{t_{c}(W)}{t})\right]
^{2}(1+28(W/U)^{2}).
\end{equation}
Therefore, as any compressibility, $\chi^{(2)}$ is found to vanish in the
vicinity of the Mott transition, even in presence of finite disorder. As a
result, the tendency to glassy ordering is strongly suppressed at weak
disorder, where one approaches the Mott insulating state.
\begin{figure}
[ptb]
\begin{center}
\includegraphics[
height=2.3346in,
width=2.8908in
]%
{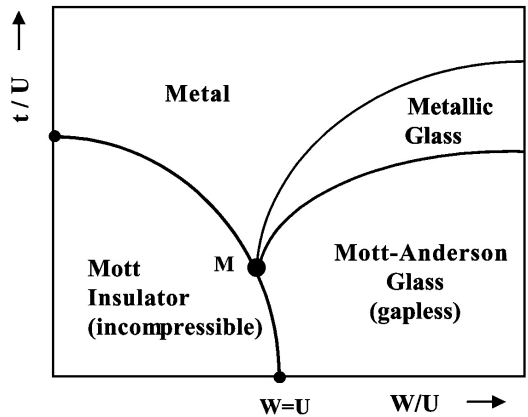}%
\caption{Global phase diagram for the disordered Hubbard model, as a function
of the hopping element $t$ and the disordered strength $W$, both expressed in
units of the on-site interaction $U$. The size of the metallic glass phase is
determined by the strength of the inter-site interaction $V$.}%
\end{center}
\end{figure}

Finally, having analyzed the limits of weak and strong disorder, we briefly
comment on what may be expected in the intermediate region $W\gtrsim U$. On
general grounds, we expect a global phase diagram as shown in Fig.2. The Mott
gap cannot exist for $W>U$, so in this region and for sufficiently small $t$
(i. e. kinetic energy), one enters an gapless (compressible) Mott-Anderson
insulator. For $W\gtrsim U,$ the computation of $\chi^{(2)}$ requires the full
solution of the Mott-Anderson problem. The required calculations can and
should be performed using the formulation of Ref. \cite{motand}, but that
difficult task is clearly beyond the scope of this letter. However, based on
general arguments presented above, we expect $\chi^{(2)}$ to \emph{vanish} as
one approaches the Mott insulator $(W<U)$, but to \emph{diverge} as one
approaches the Mott-Anderson insulator ($W>U).$ Near the tetracritical point M
(see Fig. 2), we may expect $\chi^{(2)}\sim\delta W^{-\alpha}\delta t^{\beta
},$ where $\delta W=W-W_{Mott}(t)$ is the distance to the Mott transition
line, and $\delta t=t-t_{c}(W)$ is the distance to the Mott-Anderson line.
Using this ansatz and Eq. (\ref{instab}), we find the glass transition line to
take the form%

\begin{equation}
\delta t=t_{G}(W)-t_{c}(W)\sim\delta W^{\beta/\alpha};\;W\gtrsim W_{M}.
\label{tetracritical}%
\end{equation}

We thus expect the intermediate metallic glass phase to be suppressed as the
disorder is reduced, and one approaches the Mott insulating state. Physically,
glassy behavior of electrons corresponds to many low-lying rearrangements of
the charge density; such rearrangements are energetically unfavorable close to
the (incompressible) Mott insulator, since the on-site repulsion $U$ opposes
charge fluctuations. Interestingly, very recent experiments on low density
electrons in silicon MOSFETs have revealed the existence of exactly such an
intermediate metallic glass phase in low mobility (highly disordered) samples
\cite{popovic}. In contrast, in high mobility (low disorder) samples
\cite{jan}, no intermediate metallic glass phase is seen, and glassy behavior
emerges only as one enters the insulator, consistent with our theory. Similar
conclusions have also been reported in studies of highly disordered InO$_{2}$
films \cite{zvi}, where the glassy slowing down of the electron dynamics seems
to be suppressed as the disorder is reduced and one crosses over from an
Anderson-like to a Mott-like insulator. In addition, these experiments
\cite{popovic,jan} provide striking evidence of scale-invariant dynamical
correlations inside the glass phase, consistent with the hierarchical picture
of glassy dynamics, as generally emerging from mean-field approaches
\cite{spinglass} such as the one used in this letter.

We thank S. Bogdanovich, S. Chakravarty, J. Jaroszynski, D. Popovi\'{c}, and
Z. Ovadiyahu for useful discussions. This work was supported by the NSF grant
DMR-9974311 and the National High Magnetic Field Laboratory.

\end{document}